\documentclass{jpsj3}
\usepackage{graphicx, color}
\usepackage{titlesec}

\titleformat{\subsection}{}{\thesubsection}{1em}{}

\newcommand{\dz}{$d_{z^2}$}
\newcommand{\dxz}{$d_{xz}$}
\newcommand{\dyz}{$d_{yz}$}
\newcommand{\dxzyz}{$d_{xz/yz}$}
\newcommand{\dxy}{$d_{xy}$}

\newcommand{\Ef}{$E_\mathrm{F}$}

\newcommand{\Tc}{$T_{\mathrm{c}}$}
\newcommand{\kf}{$k_{\mathrm{F}}$}

\newcommand{\doc}{$^{\circ}$C}

\newcommand{\BaFeCoAs}{$\mathrm{Ba(Fe}_{1-x}\mathrm{Co}_{x}\mathrm{)_{2}As_{2}}$}

\newcommand{\BaFeAsP}{$\mathrm{BaFe_{2}}(\mathrm{As}_{1-x}\mathrm{P}_{x}\mathrm{)_{2}}$}
\newcommand{\BaFeAsPa}{$\mathrm{BaFe_{2}}(\mathrm{As}_{0.62}\mathrm{P}_{0.38}\mathrm{)_{2}}$}

\newcommand{\SrFeAsP}{$\mathrm{SrFe_{2}}(\mathrm{As}_{1-x}\mathrm{P}_{x}\mathrm{)_{2}}$}
\newcommand{\SrFeAsPa}{$\mathrm{SrFe_{2}}(\mathrm{As}_{0.65}\mathrm{P}_{0.35}\mathrm{)_{2}}$}

\newcommand{\SrCaFeAsP}{$\mathrm{Sr}_{1-y}\mathrm{Ca}_{y}\mathrm{Fe_{2}}(\mathrm{As}_{1-x}\mathrm{P}_{x}\mathrm{)_{2}}$}

\newcommand{\Caa}{$\mathrm{Sr}_{0.92}\mathrm{Ca}_{0.08}\mathrm{Fe_{2}}(\mathrm{As}_{1-x}\mathrm{P}_{x}\mathrm{)_{2}}$}
\newcommand{\Cab}{$\mathrm{Sr}_{0.92}\mathrm{Ca}_{0.08}\mathrm{Fe_{2}}(\mathrm{As}_{0.75}\mathrm{P}_{0.25}\mathrm{)_{2}}$}

\newcommand{\etal}{\textit{et~al.}}

\title{Electronic Structure of \SrCaFeAsP\ ($x$ = 0.25, $y$ = 0.08) 
 Revealed by Angle-Resolved Photoemission Spectroscopy}

\author{Toru Adachi$^1$, Shinichiro Ideta$^2$, Zi How Tin$^1$, Hidetomo Usui$^1$, Kiyohisa Tanaka$^2$, Shigeki Miyasaka$^1$, Setsuko Tajima$^1$}
\inst{$^1$Department of Physics, Graduate School of Science, Osaka University, 
Toyonaka, Osaka 560-0043, Japan\\
$^2$UVSOR Synchrotron Facility, Institute for Molecular Science, Okazaki 444-8585, Japan}

\date{\today}

\abst{
We have investigated the electronic structure of \SrCaFeAsP\ ($x$ = 0.25, $y$ = 0.08) by means of angle-resolved photoemission spectroscopy. From the comparison with the results of \BaFeAsP, the effects of smaller structural anisotropy ($c/a$) on the Fermi surfaces (FSs) and the gap structures are discussed. The observed FSs have three dimensional shapes. One of the hole FSs is strongly warped between the $\Gamma$ and $Z$ points, and the innermost FS observed at the $Z$ point disappears at the $\Gamma$ point, which is similar to the FS features of \SrFeAsP\ ($x$ = 0.35). In the superconducting state, the node like gap-minimum is present for the \dxy\ electron FS near the $X$ point, while the gaps around the other high symmetry points are isotropic. Several theoretical models based on the spin and/or the orbital fluctuation are examined to explain all these experimental results.
}

\begin{document}

\maketitle

\section{Introduction}
The recent discovery of the iron based superconductor \cite{Kamihara} has encouraged researchers to seek a different path to the high-temperature superconductivity from that of cuprates. Iron based superconductors are characterized by multiple degrees of freedom such as spin, orbital/charge, and lattice. They are highly entangled and it is very difficult to specify which contribution is dominant for superconductivity. Depending on the pairing mechanism, a superconducting (SC) gap has a different structure. Therefore, SC gap measurements by angle-resolved photoemission spectroscopy (ARPES) could be a key experiment to determine the paring mechanism \cite{Kondo, Borisenko, Liu, Evtushinsky}. 

Among various iron based superconductors, optimally doped \BaFeAsP\ (Ba122P) is one of the candidates for a nodal superconductor. In spite of extensive experimental and theoretical efforts, however, the positions of nodes have not been determined yet \cite{Zhang, gap_Suzuki, Shimojima_SCgap, gap_Y, khodas,saito, Yamashita}. One of the ARPES measurements has pointed out that the node is on the outermost hole Fermi surface (FS) around the $Z$ point \cite{Zhang}, which can be explained by the spin-fluctuation-mediated paring mechanism \cite{gap_Suzuki}. On the other hand, the other studies have shown that the SC gaps on the hole FSs are almost momentum-independent (no node around the $Z$ point) \cite{Shimojima_SCgap} and the minimum gap exists on one of the electron FSs \cite{gap_Y}. The existence of loop nodes on the electron FSs has been predicted by the spin fluctuation theory \cite{khodas} and the spin+orbital fluctuation theory \cite{saito}. The study of angle-resolved thermal conductivity also supports the presence of loop nodes on the electron FSs \cite{Yamashita}. 

The above controversy may be caused by doping or impurity scattering dependence of fluctuation in Ba122P. In all the previous studies, optimally doped samples ($x = 0.30-0.35$, \Tc\ $\sim 30$ K) were used for ARPES measurements \cite{Zhang, Shimojima_SCgap, gap_Y}. However, since the \Tc\ of \BaFeAsP\ around the optimal doping is not so sensitive to $x$, a slight difference in $x$ may cause the great difference in fluctuation around the optimal doping (i.e. near the quantum critical point (QCP)), resulting in the discrepancy of nodal positions. 

 In order to explore this problem, more systematic studies are necessary. One of the approaches is to modify the FSs by replacing Ba into Sr or Ca, and to investigate the effects of this modification on the pairing interaction via the gap structures. Recently the ARPES study has revealed that the FSs of optimally doped \SrFeAsP\ (Sr122P) have different topology from that of Ba122P \cite{FS_Y, FS_S}. In Sr122P, one of the hole FSs shrinks from the $Z$ point to the $\Gamma$ point and finally vanishes around the $\Gamma$ point, while the corresponding FS of Ba122P remains. This difference of FSs should affect the nesting condition and thus the spin fluctuation. Therefore, the comparison of the gap structures of Ba122P and Sr/Ca122P is expected to give a useful information about the pairing mechanism of this material.  

In this study, we have investigated the electronic structure of optimally doped \Caa\ with $x=0.25$ (SrCa122P) in order to clarify the SC gap structure. The lattice anisotropy ratio of $c$- and $a$-axis ($c/a$) of this material is smaller than those of Ba122P and Sr122P, which leads to a relatively three dimensional FS. Thus, the spin fluctuation of SrCa122P should be weaker than that of Ba122P, while \Tc\ ( = 32 K) is almost the same as those of the other phosphorus substituted 122 systems \cite{Adachi}. 
By ARPES measurements using synchrotron radiation light, we found that the FSs of SrCa122P have three dimensional shapes. One of the hole FSs is strongly warped between the $\Gamma$ and $Z$ points, and the innermost FS observed at the $Z$ point disappears at the $\Gamma$ point, which is similar to the FS features of Sr122P. Moreover, the nodal behavior of the SC gap is observed on the outer \dxy\ electron FS at the $X$ point. This is different from the behavior of Ba122P for which the SC gap anisotropy is observed on the inner \dxzyz\ electron FS. The difference of nodal positions indicates that the structural anisotropy causes the change of the SC gap topology. The observed results suggest some contributions of the orbital fluctuation to the superconductivity in this system.

\section{Experiments}
Single crystals of SrCa122P were synthesized by the self-flux method as described elsewhere \cite{Adachi}. The crystals were annealed at 500 \doc\ for 4 days in order to minimize disorder effect which broaden ARPES spectra and affects the determination of a SC gap values \cite{Evtushinsky}. The highest \Tc\ value in this system (\Tc\ = 32 K with $\Delta$\Tc\ $\sim$ 3 K) was confirmed by the magnetic susceptibility measurement. ARPES measurements were carried out at BL 7U and BL 5U in UVSOR-III Synchrotron using incident photons with $h\nu=$ 14 eV to 40 eV and 40 eV to 80 eV, respectively. Energy resolution was set at $7-8$ meV and angular resolution was $\sim0.2 ^{\circ}$. Such resolution is not considered to affect the SC gap structure so much according to the previous study \cite{Evtushinsky}. Each data set is obtained in almost 6 hours and we have confirmed that the spectrum of the last acquisition did not change from the one just after cleaving. The incident photons polarized linearly and MBS A1 analyzer were used for all measurements. All samples were cleaved $in$-$situ$ at 12 K in an ultrahigh vacuum $\sim6\times10^{-9}$ Pa. Calibration of the Fermi level (\Ef) was achieved by referring to the spectra of gold which is contacted with a sample electrically. The inner potential of $V_0$ = 13.5 eV was used in order to get a best periodicity of hole FSs along the $k_z$ direction.

Note that all the ARPES spectra in the measurement are affected by the $k_z$ broadening effect due to the short escape depth($\lambda$) of photoelectrons. The escape depth of photoelectrons for the incident photons with $h\nu=$ 40 eV is $\sim$ 4 $\mathrm{\AA}$ and $1/\lambda$ is approximately $0.4\times (2\pi/c)$. ARPES spectra are broadended by the Lorentzian function with this width in the $k_z$ direction and this may affect the estimation of a FS shape or a SC gap.

The calculation for the band structure of SrCa122P was performed using the projector-augmented-wave method \cite{band1} and the Perdew-Burke-Ernzerhof parameterization of the generalized-gradient approximation \cite{band2} (PBE-GGA). They are implemented in the Vienna $ab$ $initio$ simulation package \cite{band3,band4,band5,band6}. $10\times 10\times 10$ $k$-point meshes and 800 eV of cut-off energy were taken. We used the virtual crystal approximation to take into account the effect of partial substitution of Sr by Ca and As by P, respectively. In order to obtain the orbital component of the band structure, a 10-orbital tight-binding Hamiltonian was constructed by exploiting the maximally localized Wannier functions \cite{band7,band8,band9} mainly originating from the five 3$d$ orbitals of Fe. For this calculation, the structure parameters that were experimentally determined for the present crystal were used \cite{Adachi}. 

\section{Results}

\subsection{Electronic structure in the normal state}
\begin{figure*}[htpb]
  \centering
  \includegraphics[width=150mm,origin=c,keepaspectratio,clip]{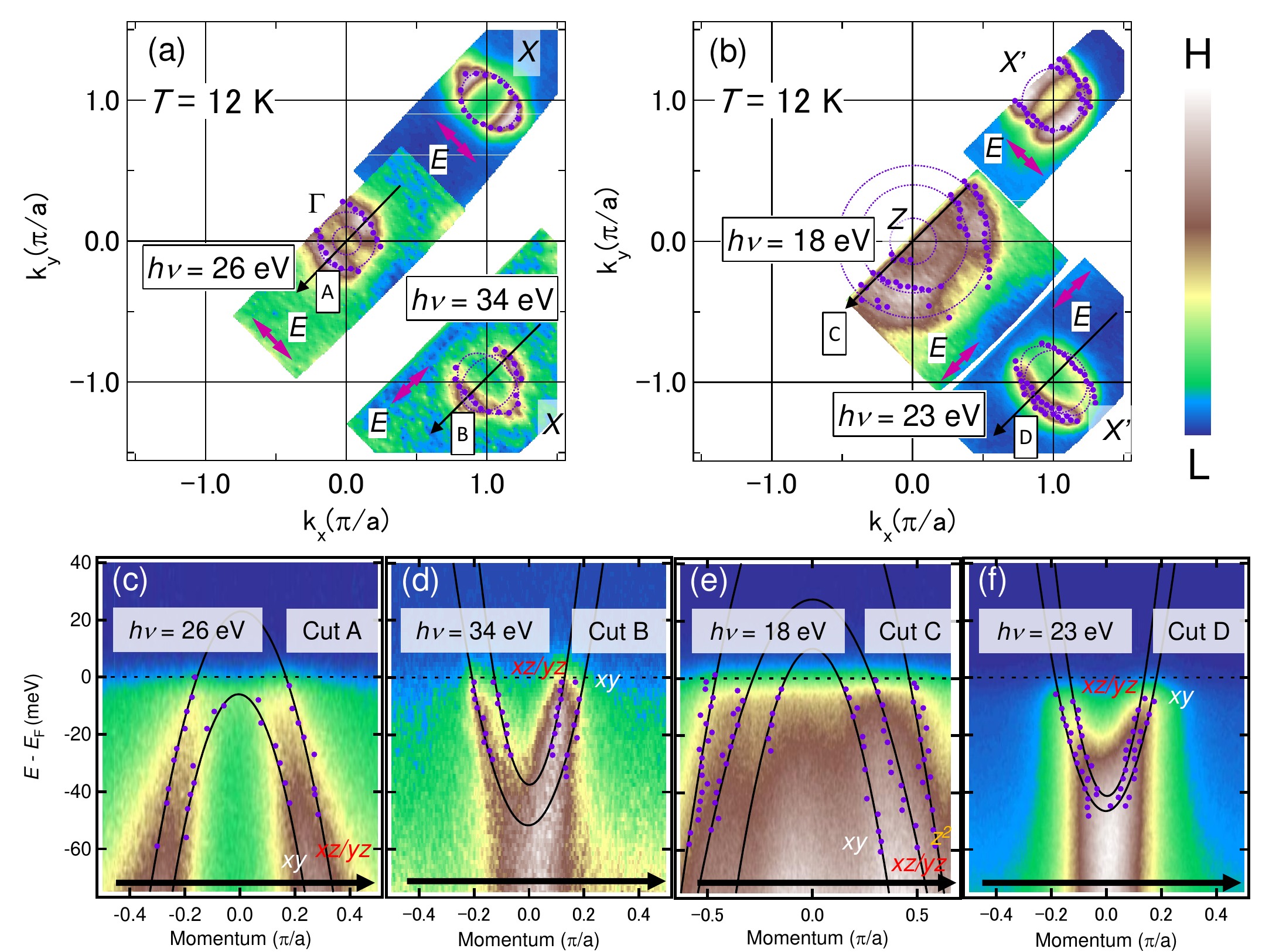}
  \caption{(Color online) Electronic structure of \Cab. (a),(b) Intensity map at \Ef\ in the $k_x-k_y$ plane at $k_z \sim 0$ and $2\pi/c$. (c)-(f) Energy-momentum ($E$-$k$) plots corresponding to the cuts in (a) and (b). Purple dots indicate the peak positions estimated from momentum-distribution curves (MDCs) at several energies and are fitted with parabolic functions shown as black curves.}
  \label{fig1}
\end{figure*}

Figure \ref{fig1} shows the FS mappings and energy-momentum ($E-k$) plots of SrCa122P observed by ARPES in the $k_x-k_y$ plane at $k_z \sim 0$ ($\Gamma$) and $2\pi/c$ ($Z$). We chose different photon energies for measurements of hole and electron FSs in order to observe both FSs on the same $k_z$ plane. As seen in Figs. \ref{fig1}(a) and (b), there are observed two hole FSs around the $\Gamma$ point, three hole FSs around the $Z$ point, and two electron FSs around the $X$ point. Note that only one hole FS around the $\Gamma$ point has been observed using $s$-polarized light, while the other inner hole FS can be observed using $p$-polarized light as shown later in Fig. \ref{fig5}. The energy position of the inner hole band tops is below \Ef\ around the $\Gamma$ point as shown in Fig. \ref{fig1}(c). One can clearly see that the shape of the electron FS at $k_z \sim 0$ is very similar to that of the electron FS at $k_z \sim 2\pi/c$ if it is rotated by 90 deg \cite{I4mmm, sunagawa}. 

The $E-k$ plots are indicated in Figs. \ref{fig1}(c)-\ref{fig1}(f), where the data points were determined from the momentum distribution curves (MDCs) at several energies \cite{Supplement}. All the data points are well fitted with parabolic functions. Although some of the fitting curves seems optimistic, we have confirmed its validity from the second derivative of intensity \cite{Supplement}. Here we note that the energy position of the inner hole band top is below \Ef\ around the $\Gamma$ point as shown in Fig. \ref{fig1}(c).
The band dispersions and the orbital characters of all FSs are described in Figs. \ref{fig1}(c)-\ref{fig1}(f). Each orbital characterization was achieved by considering polarization dependence of the photoemission matrix element of each orbital and the intensity of each band, which is consistent with the band calculation as described later. 

\begin{figure*}[htpb]
  \centering
  \includegraphics[width=150mm,origin=c,keepaspectratio,clip]{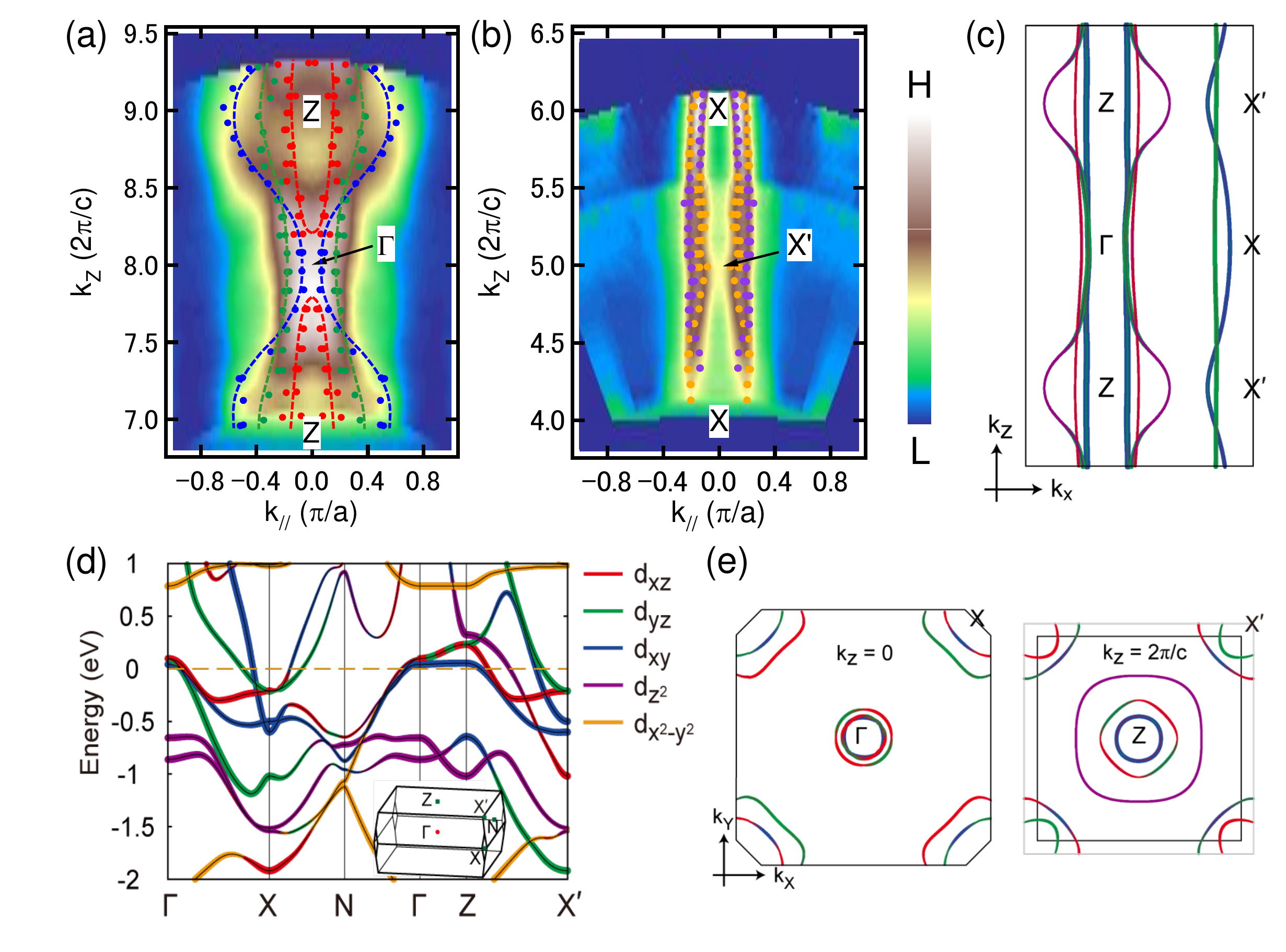}
  \caption{(Color online) FS mapping in $k_{//}-k_z$ plane for \Cab. (a) Hole FSs and (b) electron FSs observed with $p$-polarized light. $k_{//}$ is along the $\Gamma$-$X$ line. (c)-(e) Calculated FSs and band dispersions for \Cab.}
  \label{fig2}
\end{figure*}

Figures \ref{fig2}(a) and \ref{fig2}(b) illustrate the FS mappings of the hole and electron FSs in the $k_{//}-k_z$ plane, respectively. We assign the inner (red), middle (green), and outer (blue) hole FSs around the $Z$ point to the $\alpha$, $\beta$, and $\gamma$ FSs and two electron FSs to the $\delta$ and $\epsilon$ FSs, respectively. Hole FSs show the strong $k_z$-dependence, while electron FSs show weak modulation, namely, relatively two dimensional. This trend is consistent with the first principle band calculation displayed in Fig. \ref{fig2}(c). The orbital characters derived from the calculation are consistent with the polarization dependence of each band. The relatively weak $k_z$ dependence of electron FSs is reproduced by the band calculation, whereas the in-plane FS shapes are different in the experimental result (Fig. \ref{fig1}) and the calculation in Fig. \ref{fig2}(e); the innermost \dxy\ hole FS. 

In the calculation, the top of the \dxy\ hole band is above \Ef\ in the entire $k_z$ range, while we have observed this band below \Ef\ around the $\Gamma$ point in the present ARPES measurement. The discrepancy between the experiment and the calculation indicates constraints on the present band calculation that something important is not taken into account. 

\begin{figure}[htbp]
  \centering
  \includegraphics[width=85mm,clip]{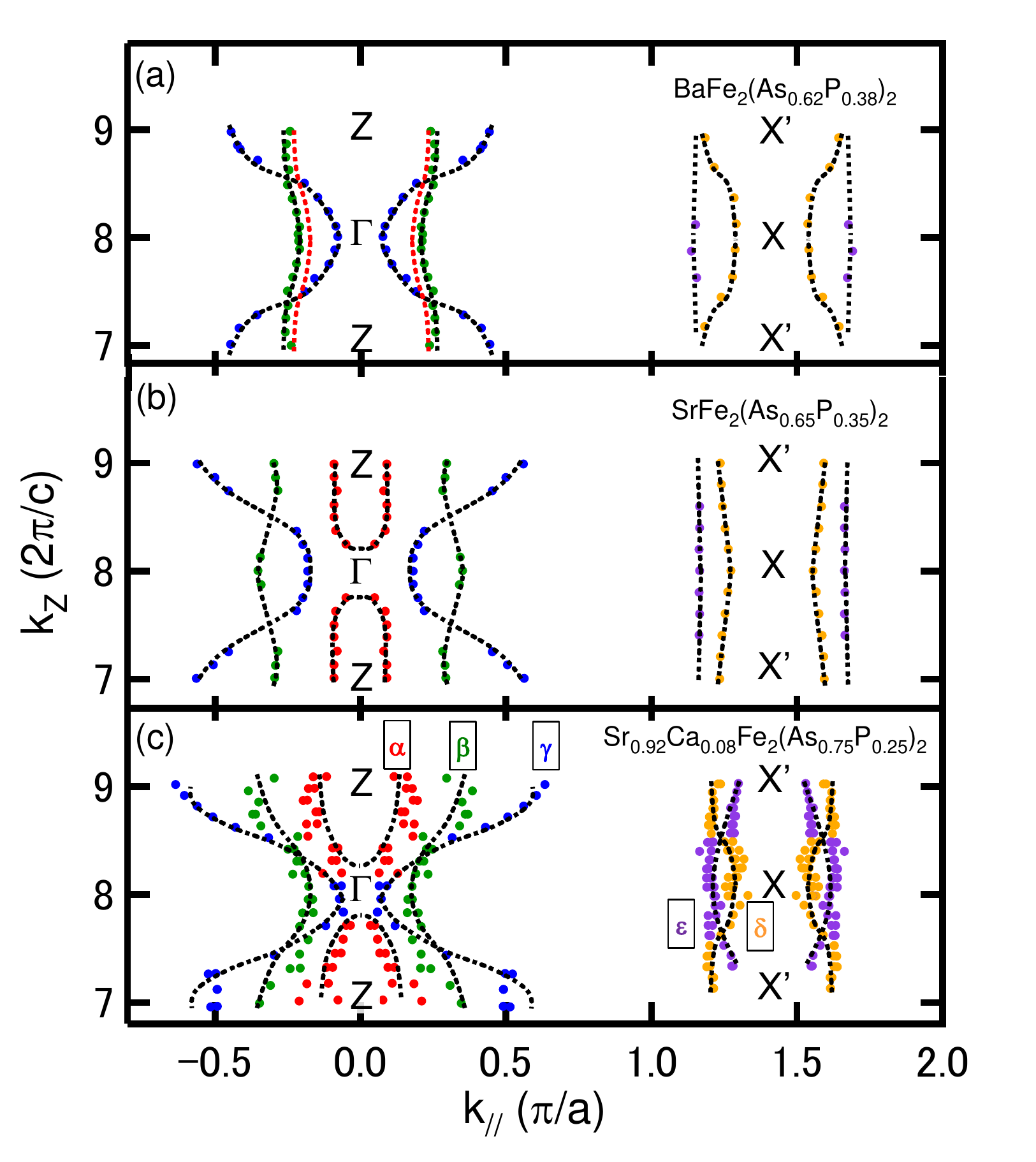}
    \vspace{0.5cm}
  \caption{(Color online) FSs in the $k_{//}-k_z$ plane for (a)\BaFeAsPa, (b)\SrFeAsPa, and (c)\Cab. Data of \BaFeAsPa\ and \SrFeAsPa\ were taken from Refs. \cite{FS_Y, FS_S} The red broken line of \BaFeAsPa\ is determined by the data of Ba122P \cite{gap_Y}. The red, green, and blue dots indicate the $\alpha, \beta,$ and $\gamma$ hole FSs, respectively and orange and purple dots indicate the electron FSs.}
  \label{fig3}
\end{figure}

To demonstrate the effect of structural anisotropy ($c/a$), we compare the FSs of SrCa122P in the $k_{//}-k_z$ plane with those of \BaFeAsPa\ and \SrFeAsPa\ in Fig. \ref{fig3} \cite{FS_Y, gap_Y, FS_S}. From $A$ = Ba to SrCa, the $A$ ion radius in $A$122P and thus the unit cell volume become smaller. Optimally doped crystals of these compounds have almost the same shaped FeAs$_4$ tetrahedra and then the difference of the FS topology seen in Fig. \ref{fig3} can be attributed to the $c/a$ structural anisotropy. One can easily see the big difference in the hole FSs while the $k_z$ dependences of the electron FSs are nearly the same. As the $A$ ionic radius becomes smaller, the \dz\ outermost $\gamma$ FS (indicated by blue dots) around the $Z$ point becomes bigger and the \dxzyz\ inner $\gamma$ FS around the $\Gamma$ point becomes smaller. On the other hand, the \dxzyz\ $\beta$ FS (green) is only moderately deformed. The $\alpha$ FS (red) is considered to be of almost the same size and shape with the $\beta$ FS in Ba122P \cite{gap_Y}. However, the $\alpha$ FS shows a remarkable $c/a$ dependence, and disappears around the $\Gamma$ point in \SrFeAsPa\ and SrCa122P, while it remains in the entire $k_z$ Brillouin zone in \BaFeAsPa. One may notice that the $\beta$ FS is convex around $\Gamma$ in \SrFeAsPa\ while it is concave in \BaFeAsPa\ and SrCa122P. The origin of this difference is unclear.



It is interesting to examine the effective masses because a remarkable mass enhancement near QCP has been reported in the de Haas-van Alphen measurement of Ba122P \cite{dhva}. Combining the ARPES data and the band calculation, we can evaluate the effective mass of each band around the high symmetry point as summarized in Table \ref{mass}, where $m^*$, $m_e$, and $m_b$ are the effective mass, the free-electron mass, and the band mass, respectively. The effective masses were obtained by fitting the bands in Figs. \ref{fig1}(c)-\ref{fig1}(f) and \ref{fig5}(f) with parabolic functions. The band masses were deduced in the same manner using the calculated bands in Fig. \ref{fig2}(d). One can see that the effective mass of electron bands are lighter than those of hole bands. Interestingly, the strong mass enhancement has been observed for one of the electron FSs, as reported in the previous ARPES measurements for Ba122P and Sr122P \cite{FS_Y, FS_S}, implying the same mechanism for this quantum critical behavior. The strong mass enhancement is also seen in the $\beta$ hole band around the $Z$ point. These mass enhancements of $\beta$ (\dxzyz) and $\epsilon$ (\dxy) bands imply the strong interorbital scattering.

\begin{table}[htb]
\caption{Effective masses and band masses of \SrCaFeAsP\ ($x$= 0.25, $y$ = 0.08). $m^*, m_e$, and $m_b$ are the effective mass, free-electron mass, and band mass, respectively. All values are those along the cuts in Figs. \ref{fig1}(c)-\ref{fig1}(f).}
\label{mass}
\begin{center}
\begin{tabular}{lcccccc} \hline
\hline
High symmetry point & FS & orbital& $m^*/m_e$ & $m_b/m_e$ & $m^*/m_b$ \\ 
\hline
$\Gamma$& $\beta$ & \dxz & 2.8 & 1.5 & 1.9 \\ 
 & $\gamma$ & \dyz & 2.4 & 0.9 & 2.8 \\ 
$Z$ & $\alpha$ & \dxy & 3.9 & 1.8 & 2.1 \\ 
 & $\beta$ & \dxz & 7.2 & 1.2 & 6.2 \\
 & $\gamma$ & \dz & 6.0 & 2.7 & 2.2 \\ 
$X$ & $\delta$ & \dxy & 1.9 & 1.0 & 1.8 \\ 
 & $\epsilon$ & \dxzyz & 1.3 & 1.6 & 0.8 \\ 
$X'$ & $\delta$ & \dxzyz & 1.1 & 1.9 & 0.6 \\ 
 & $\epsilon$ & \dxy & 1.8 & 0.3 & 6.4 \\ 

 \hline
 \hline
\end{tabular}
\end{center}
\end{table}

\subsection{Superconducting gap}
\begin{figure*}[htpb]
  \centering
  \includegraphics[width=150mm,origin=c,keepaspectratio,clip]{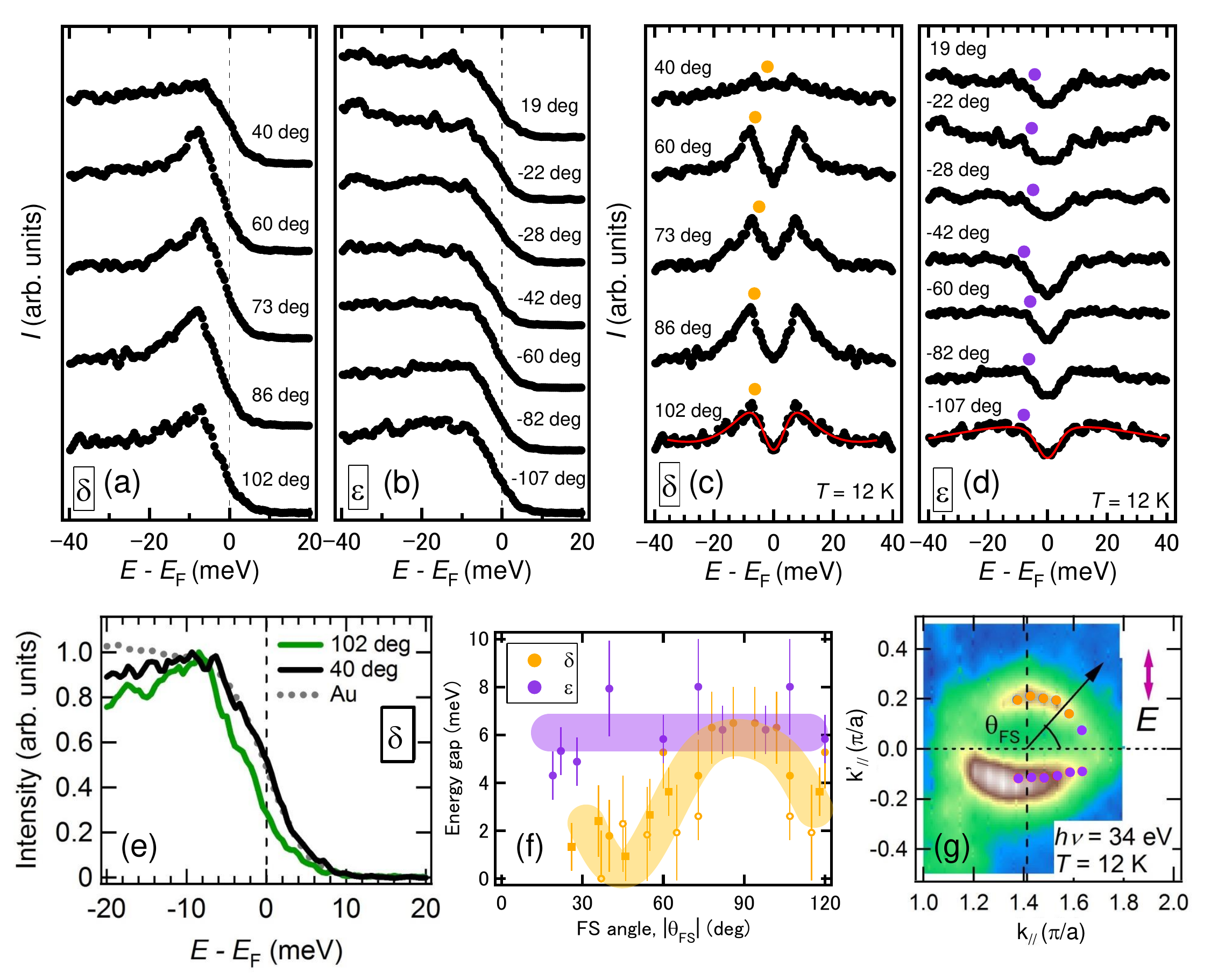}
  \caption{(Color online) SC gaps for \Cab\ around the $X$ point taken at $h\nu = 34$ eV on the $\delta$ and $\epsilon$ FSs. Energy-distribution curves (EDCs) at Fermi momentum (\kf) on the $\delta$ and $\epsilon$ are shown in (a) and (b), respectively. EDCs symmetrized at \Ef\ for $\delta$ and $\epsilon$ FSs are shown in (c) and (d), respectively. (e) EDCs for the $\delta$ FS at $\theta_\mathrm{FS}$ = 40, 102 deg, and the spectrum of gold. (f) Energy gaps plotted as a function of the FS angle, $\theta_\mathrm{FS}$. As for the $\delta$ FS, SC gaps observed from other two samples are also plotted (open circles and squares). All data are symmetrized by taking into account the space group symmetry. (g) In-plane FS mapping around the $X$ point. $k_{//}$ and $k'_{//}$ are along $\Gamma$-$X$ lines.}
  \label{fig4}
\end{figure*}

Next, we present the SC gap structure around various high symmetry points for SrCa122P. 
The results around the $X$ point ($\pi/a, \pi/a, 0$) are shown in Fig. \ref{fig4}. We have observed two electron FSs, \dxy\ ($\delta$) and \dxzyz\ ($\epsilon$), around the $X$ point. The raw EDCs in Figs. \ref{fig4}(a) and \ref{fig4}(b) were symmetrized at \Ef\ and re-plotted in Figs. \ref{fig4}(c) and \ref{fig4}(d). The symmetrizing procedure is widely used to remove the broadening effect of Fermi-Dirac (FD) function convoluted with the instrumental resolution function and to determine gap values \cite{Norman}. In Fig. \ref{fig4}(c) for the $\delta$ FS, we can clearly see the peak structure that can be assigned to a SC gap. Although no clear peak is seen in Fig. \ref{fig4}(d) ($\epsilon$ FS), the shoulder structure (the edge of V-shape) is evident. Comparing the normal state spectrum $I(37 \mathrm{K})$ and the SC state one $I(12 \mathrm{K})$ for the electron FSs around the $X'$ point, we plotted the divided spectra $I(37 \mathrm{K})/I(12 \mathrm{K})$ and found that they showed clear peaks corresponding to the SC gaps \cite{Supplement}. These peak energies well agree with the shoulder energies in the symmetrized spectra, which supports that we are allowed to assign the shoulder structures to the SC gaps. The results in Figs. \ref{fig4}(c) and \ref{fig4}(d) have revealed that the $\delta$ electron FS has a gap minimum indicating the existence of the node, while the $\epsilon$ electron FS has an almost isotropic gap. 

To demonstrate this gap minimum more explicitly, we replot the spectra of this sample at $\theta_\mathrm{FS}$ = 40 deg and 102 deg together with the spectrum of gold in Fig. \ref{fig4}(e). It is shown that the spectral shape at $\theta_\mathrm{FS}$ = 40 deg is very close to that of gold, namely, the gap closes, while the leading edge at $\theta_\mathrm{FS}$ = 102 deg is shifted and thus the gap obviously opens. 

In order to derive the SC gap size quantitatively, the symmetrized EDCs have been fitted with the well-known phenomenological SC spectral function proposed by M. R. Norman $et$ $al.$ \cite{Norman2} The detail of the fitting procedure can be found in the Supplemental Material \cite{Supplement}. Note that some colored dots in Figs. \ref{fig4}(c) and (d) do not coincide well with the peak positions of the EDCs because of the finite energy resolution and the linear background \cite{Supplement}. All fitting results are summarized in Fig. \ref{fig4}(f) which includes the results of the other two samples for $\delta$ electron FS (open circles and squares). One can see that the $\delta$ electron FS clearly shows the existence of the gap minimum around $\theta_\mathrm{FS}$ = 45 deg. 

\begin{figure*}[htpb]
  \centering
  \includegraphics[width=150mm,origin=c,keepaspectratio,clip]{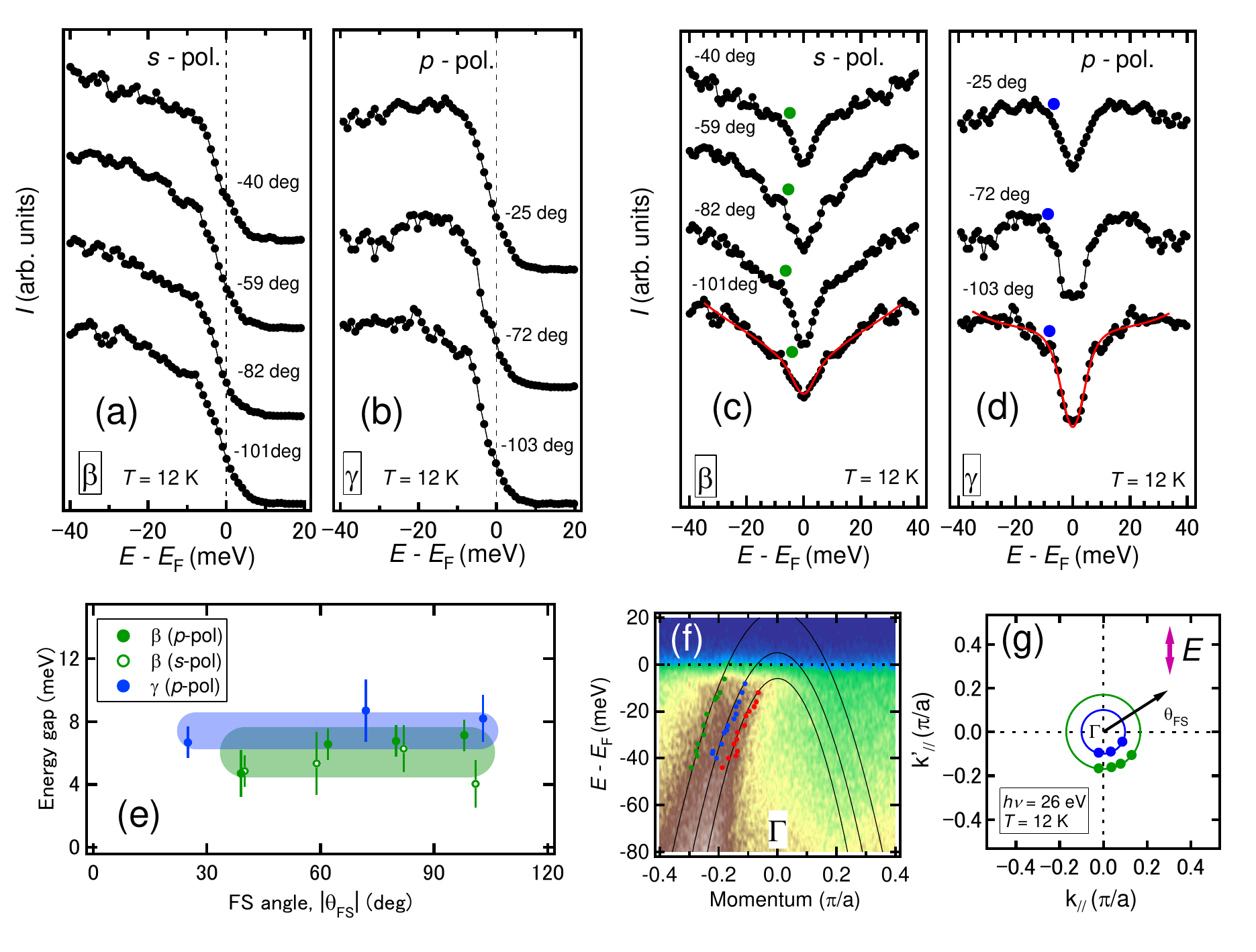}
  \caption{(Color online) SC gaps for \Cab\ around the $\Gamma$ point taken at $h\nu = 26$ eV on the $\beta$ and $\gamma$ FSs. EDCs at \kf\ on the $\beta$ and $\gamma$ FSs are shown in (a) and (b), respectively. EDCs symmetrized at \Ef\ for $\beta$ and $\gamma$ FSs are shown in (c) and (d), respectively. (e) Energy gaps plotted as a function of FS angle, $\theta_\mathrm{FS}$. (f) $E-k$ plot around the $\Gamma$ point. Green, blue, and red dots indicate peak positions estimated from MDCs and fitted by parabolic function. (g) Schematic FSs at the $\Gamma$ point and \kf -positions for the present gap measurements. $k_{//}$ and $k'_{//}$ are along $\Gamma$-$X$ lines.}
  \label{fig5}
\end{figure*}

Compared to the results around the $X$ point, the gap features are weaker around $\Gamma$ and $Z$ points. Figures \ref{fig5}(a) and \ref{fig5}(b) illustrate EDCs at Fermi momentum (\kf) for the $\beta$ and $\gamma$ bands around the $\Gamma$ point, using $s$-, and $p$-polarized light, respectively. The \kf\ positions are indicated by markers in Fig. \ref{fig5}(g). Figures \ref{fig5}(c) and \ref{fig5}(d) are the symmetrized EDCs for the $\beta$ and $\gamma$ FSs. Although no coherence peak is seen, the shoulder features can be recognized. As mentioned above, we estimate gap values from the fitting process using the phenomenological SC spectral function. It is worth noting here that the broad hump structures between $10-20$ meV in Fig. \ref{fig5}(d) reflect the effect of the inner $\alpha$ band (below \Ef) indicated by red dots in Fig. \ref{fig5}(f). The estimated gap values are plotted with respect to the FS angle, $\theta_\mathrm{FS}$ in Fig. \ref{fig5}(e). From this result, it is concluded that an almost isotropic gap opens around the $\Gamma$ point. The SC gap were estimated as $5.7\pm 0.5$ meV and $7.8\pm 0.9$ meV for $\beta$ and $\gamma$ FSs, respectively. 

\begin{figure*}[htpb]
  \centering
  \includegraphics[width=110mm,origin=c,keepaspectratio,clip]{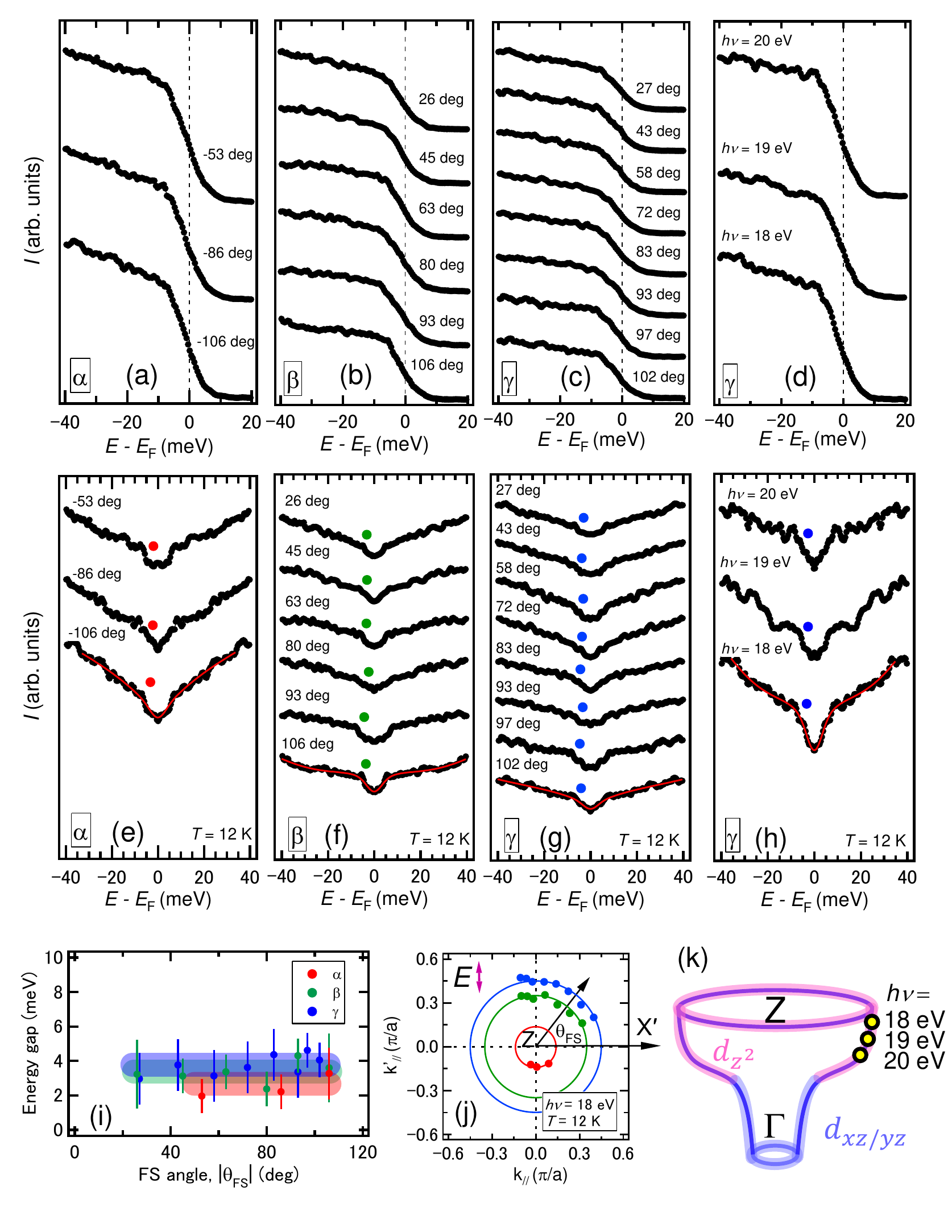}
  \caption{(Color online) SC gaps for \Cab\ around the $Z$ point on the $\alpha$, $\beta$, and $\gamma$ FSs. EDCs at \kf\ taken at $h\nu = 18$ eV on the $\alpha$, $\beta$, and $\gamma$ FSs are shown in (a)-(c), respectively. EDCs symmetrized at \Ef\ for the $\alpha$, $\beta$, and $\gamma$ FSs are shown in (e)-(g), respectively. The energy gaps estimated from (e)-(g) are plotted as a function of the FS angle, $\theta_\mathrm{FS}$ in (i). (j) Schematic FSs around the $Z$ point. $k_{//}$ and $k'_{//}$ are along $Z$-$X'$ lines. (d), (h) Photon-energy dependence of the SC gap at $\theta_\mathrm{FS} \sim 90$ deg in the $k_z$ direction. EDCs (d) and symmetrized EDCs (h) for the $\gamma$ FS. The photon energies of 18-20eV give the $k_z$ positions shown in (k).}
  \label{fig6}
\end{figure*}

The spectra around the $Z$ point are presented in Fig. \ref{fig6}. 
The EDCs for the $\alpha, \beta$, and $\gamma$ FSs are shown in Figs. \ref{fig6}(a)-(c), respectively, while the corresponding EDCs symmetrized at \Ef\ are plotted in Figs. \ref{fig6}(e)-(g). Although the gap features are very weak also around the $Z$ point, we estimated the gap values through the fitting procedure with symmetrized EDCs. The average gap magnitudes are $2.5\pm 0.7, 3.3\pm 0.6$, and $3.7\pm 0.6$ meV for $\alpha, \beta$, and $\gamma$ FSs, respectively.
As plotted in Fig. \ref{fig6}(i), the SC gaps for the $\alpha$, $\beta$, and $\gamma$ hole FSs around the $Z$ point are nearly isotropic within the error bars. 
We also examine the $k_z$-dependence of the SC gap anisotropy which is predicted by the spin fluctuation theory \cite{gap_Suzuki}. We can sweep the $k_z$ position by changing the incident photon energy, as illustrated in Fig. \ref{fig6}(k). Figures \ref{fig6}(d) and \ref{fig6}(h) show the EDCs and the symmetrized EDCs on the outermost \dz\ FS at $\theta_\mathrm{FS} \sim 90$ deg around the $Z$ point for three different incident photon energies.  We could observe no gap minimum at any $k_z$ position. 

\section{Discussion}
 We have observed the nodal gap behavior for the $\delta$ electron FS (\dxy) around the $X$ point, while the gaps were isotropic around the other high symmetry points. Although the observed SC gap structure plays a pivotal role to elucidate a pairing mechanism, no theoretical study has been reported for SrCa122P yet.
 
 The theoretical prediction for the SC gap symmetry in the isostructural Ba122P was reported by Suzuki, Usui, and Kuroki \cite{gap_Suzuki}. They have calculated the SC gap function based on the intraorbital spin fluctuation and predicted the existence of a horizontal node on the $\gamma$ FS (\dz) around the $Z$ point. Although the spin fluctuation has been confirmed by a lot of experiments, the present results do not support this scenario from many viewpoints. Firstly, no distinct gap anisotropy has been observed for $\gamma$ FS around the $Z$ point, whereas the theory predicts a horizontal node at a certain $k_z$ on this FS \cite{gap_Suzuki}. Secondly, the gap size for the $\gamma$ FS along the $Z$ - $X'$ line is comparable to those for $\alpha$ and $\beta$ FSs, whereas the theory predicts a smaller gap for the $\gamma$ FS around the $Z$ point than those of the other FSs due to the absence of the nesting channel of \dz\ bands between hole and electron FSs. Thirdly, although the theory predicts neither anisotropy nor node on the electron FSs \cite{gap_Suzuki}, the present results indicate a gap minimum on one of the electron FSs around the $X$ point. All these inconsistencies suggest that some interactions other than the spin fluctuation contribute to the superconductivity in this system. 

Here, we have to remind that all these calculations based on the spin fluctuation were demonstrated for Ba122P but not for SrCa122P. If the FS topology changes, the gap values and anisotropy may also change. For example, when the \dxy\ hole FS is missing as predicted in LaFePO, the line node could appear on the electron FSs \cite{kuroki2nd}. Another theoretical model based on the spin fluctuation has proposed that the hybridization between two electron FSs changes a line node to a loop node of the SC gap on the electron FSs, assuming that the \dxy\ hole FS is absent \cite{khodas}. Therefore, to examine a possibility of the spin fluctuation model, it is highly desired to calculate the spin fluctuation strength based on the observed FSs of SrCa122P.

The next candidate is the model where the Cooper pairing is mediated by the orbital fluctuation \cite{kontani, Onari}. The orbital fluctuation is induced by the electron-phonon interaction through the Fe ion vibration or the Coulomb interaction derived from the Aslamazov-Larkin-type vertex correction. Saito, Onari, and Kontani have proposed that a nodal gap structure should be observed in P doped 122 systems due to the competition between the attractive (orbital) and the repulsive (spin) interactions \cite{saito}. According to this theoretical model, the nodal position strongly depends on the interaction strengths. The node on the $\gamma$ FS around the $Z$ point is reproduced when the Coulomb interaction $U$, a measure of the spin fluctuation, is finite and the quadrupole interaction $g$, a measure of the orbital fluctuation, is zero. With increasing $g$, the node on the $\gamma$ FS around the $Z$ point disappears and a finite gap opens, while the loop nodes appear on the electron FSs. 

In SrCa122P, the minimum of the SC gap is present on the outer \dxy\ electron FS, while in Ba122P, it was observed on the inner \dxzyz\ electron FS \cite{gap_Y}. According to the theory \cite{saito}, the node on the \dxy\ electron FS can be reproduced when $U$ for SrCa122P is smaller than that for Ba122P. Decreasing $U$ is equivalent to weakening the contribution of the spin fluctuation. In fact, the weakness of the spin fluctuation is indicated  by the previous nuclear magnetic resonance experiments \cite{dulguun, miyamoto}. From the measurement of the spin relaxation rate $T_1$, Dulguun \etal\ have argued that the AFM spin fluctuation which is enhanced upon cooling is stronger in Ba122P than in Sr122P. The weaker AFM spin fluctuation of Sr122P is attributed to the smaller $c/a$. Because the anisotropy of SrCa122P is further smaller and its FS is more three dimensional, it is expected that the AFM spin fluctuation of this compound is also weaker than that of Ba122P. 

\begin{figure}[htpb]
  \centering
  \includegraphics[width=85mm,origin=c,keepaspectratio,clip]{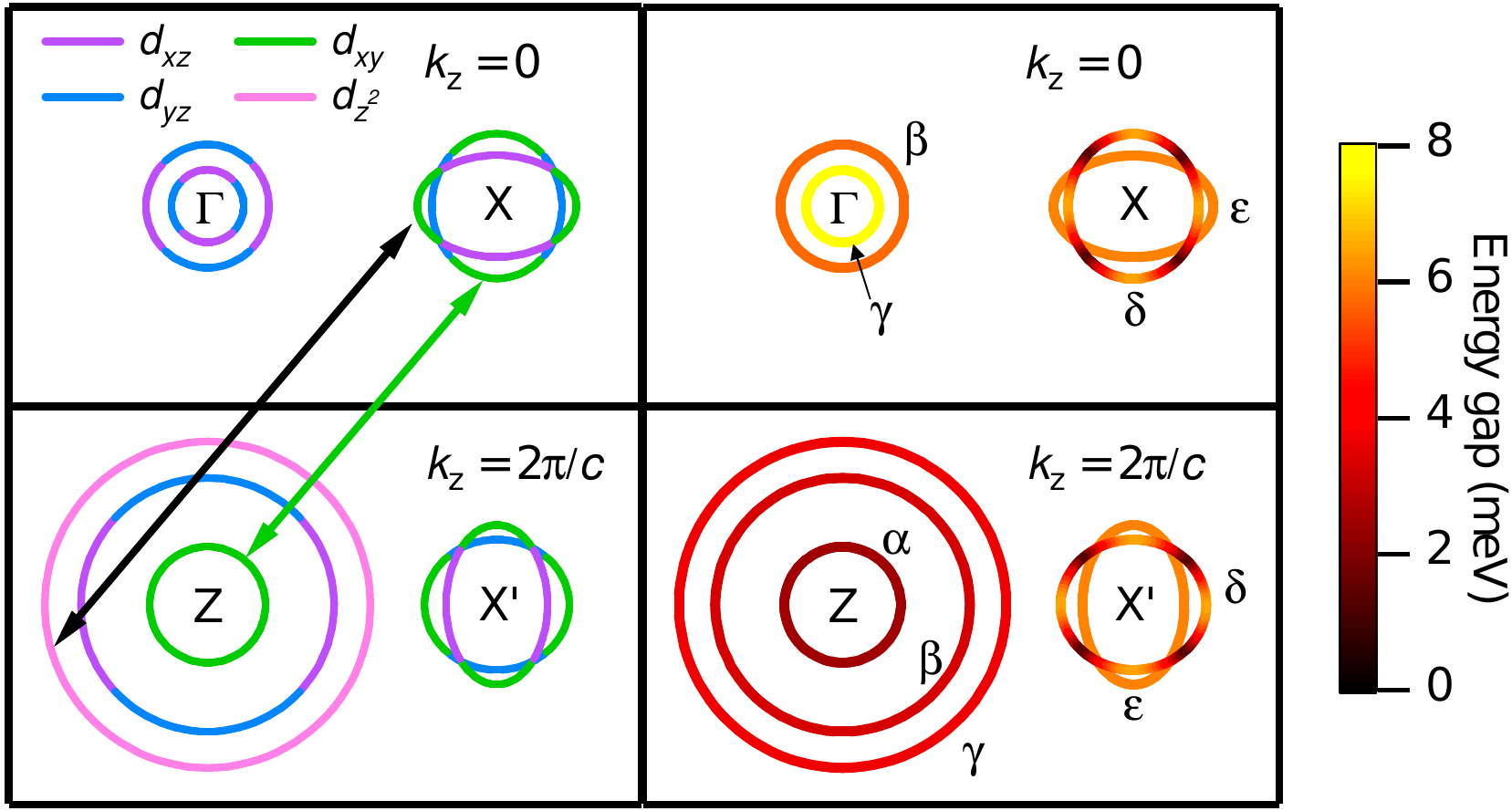}
  \caption{(Color online) Schematic FSs (Left) and color plots of the gap distribution on the FSs (Right) of \Cab. The black arrow indicates the interorbital interaction which causes a full gap on the $\gamma$ hole FS. The green arrow indicates the \dxy\ intraorbital interaction which gives rise to a node on one of the electron FSs.}
  \label{fig7}
\end{figure}

To summarize, the spin+orbital fluctuation theory qualitatively explains the present results as follows. The inter- and intra-orbital interaction yields the full gap on the $\gamma$ FS around the $Z$ point and the strong \dxy\ intraorbital ($\pi/a, \pi/a, 2\pi/c$) nesting between the hole and the electron FSs causes the sign change on the small part of the $\delta$ electron FSs as shown in Fig. \ref{fig7}. Although the orbital fluctuation indeed contributes to the SC mechanism, the FS nesting related with the spin fluctuation also plays a key role to keep the high-\Tc\ superconductivity in SrCa122P regardless of the strong three dimensionality of the hole FSs. To explain the experimental fact that the \Tc\ of P doped 122 system is insensitive to the FS dimensionality, the spin fluctuation is not enough. It is important to clarify whether the spin and the orbital fluctuations can cooperate to enhance \Tc\ or not.

\section{Conclusion}
We have observed the electronic structure of \Cab\ using ARPES and confirmed that the structural anisotropy, $c/a$, greatly changes the FS topology especially for the hole FSs. The obtained hole FSs were strongly warped in the $k_z$ direction, which is similar to those of Ba122P, but the FS size at the $Z$ point is larger than that of Ba122P. 
In the SC state, the gap minimum, indicative of a node, has been observed for the $\delta$ electron FS (\dxy) around the $X$ point. The difference of node position in Ba122P and SrCa122P suggests the weakening of spin fluctuation contribution with decreasing the structural anisotropy even if the $T_c$ does not change. The gaps were isotropic around the other high symmetry points. The present results cannot be explained by the spin fluctuation model but rather support the theory in which the orbital fluctuation plays a role in superconductivity in addition to the spin fluctuation. To settle down the problem, the specific calculation based on the anomalous FSs of SrCa122P is missed and thus highly desired. 

\begin{acknowledgments}
We thank K. Kuroki, M. Nakajima, and S. Kimura for the fruitful discussion. The ARPES experiments were carried out at UVSOR Synchrotron Facility (Proposal Nos. 27-541, 27-815, 28-532, 28-823, 28-836, 29-533, 29-552, and 29-843). This work was supported by Grants-in-Aid for Scientific Research from JSPS, Japan. T.A. acknowledges the Grant-in-Aid for JSPS Fellows.
\end{acknowledgments}

\newpage

\title{\bf{Supplemental Materials of ``Electronic Structure of \SrCaFeAsP\ ($x$ = 0.25, $y$ = 0.08) 
 Revealed by Angle-Resolved Photoemission Spectroscopy''}}

\maketitle

\section*{MDCs at various binding energies near the high symmetry points}
\begin{figure*}[htpb]
  \centering
  \includegraphics[width=150mm,origin=c,keepaspectratio,clip]{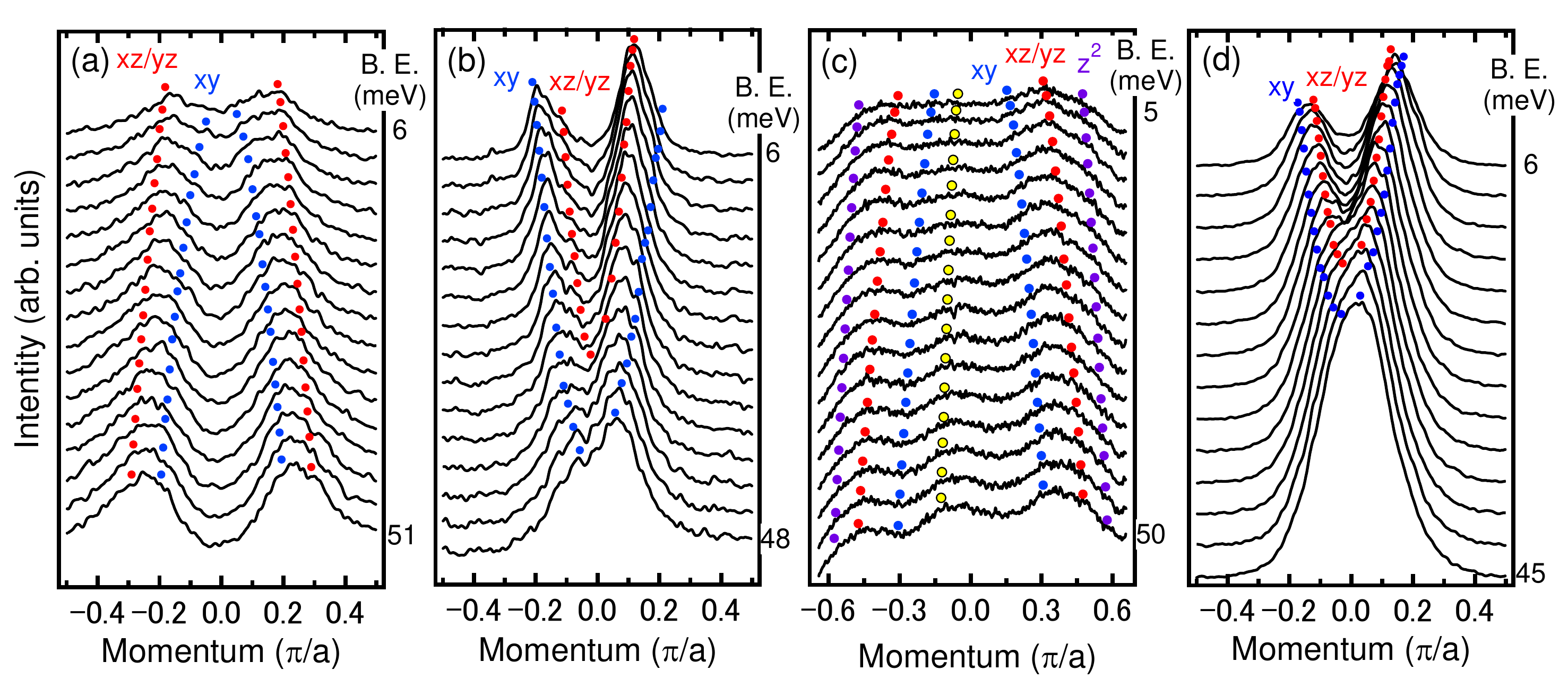}
  \caption{(Color online) MDCs at various binding energies near the high symmetry points for \Cab. MDCs in (a)-(d) are taken from the data for Figs. 1(c)-(f), respectively. Red, green, blue, orange, and purple dots indicate the hole and electron band dispersions. The yellow dots in (c) indicate the undefined band dispersion.}
  \label{figS1}
\end{figure*}

Figure \ref{figS1} shows momentum distribution curves (MDCs) at various binding energies near the $\Gamma$, $X$, $Z$, and $X'$ points. MDCs are taken from the data for Figs. 1(c)-(f). Colored dots except for yellow dots are put along the black curves in Figs. 1(c)-(f). One can see these dots are located at the peak or hump structures corresponding to the band dispersions. The hump structures at the yellow dots in Fig. \ref{figS1}(c) do not correspond to any band around the $Z$ point because there is no band dispersion around these positions with $s$-polarized light. If this band corresponds to the innermost FS and has \dxy\ orbital character, it should be more visible with $s$-polarized light. One of the possibilities is that it is a surface-related band. Indeed, the previous study has shown that the innermost surface-related band exists around the $\Gamma$ point in \BaFeCoAs\ \cite{Heumen}. Considering the very weak $k_z$ dependence of the surface state, it is not surprising to observe this band around the $Z$ point in this measurement.

\newpage
\section*{The second derivative of the intensity at $X$ point}
\begin{figure}[htpb]
  \centering
  \includegraphics[width=80mm,origin=c,keepaspectratio,clip]{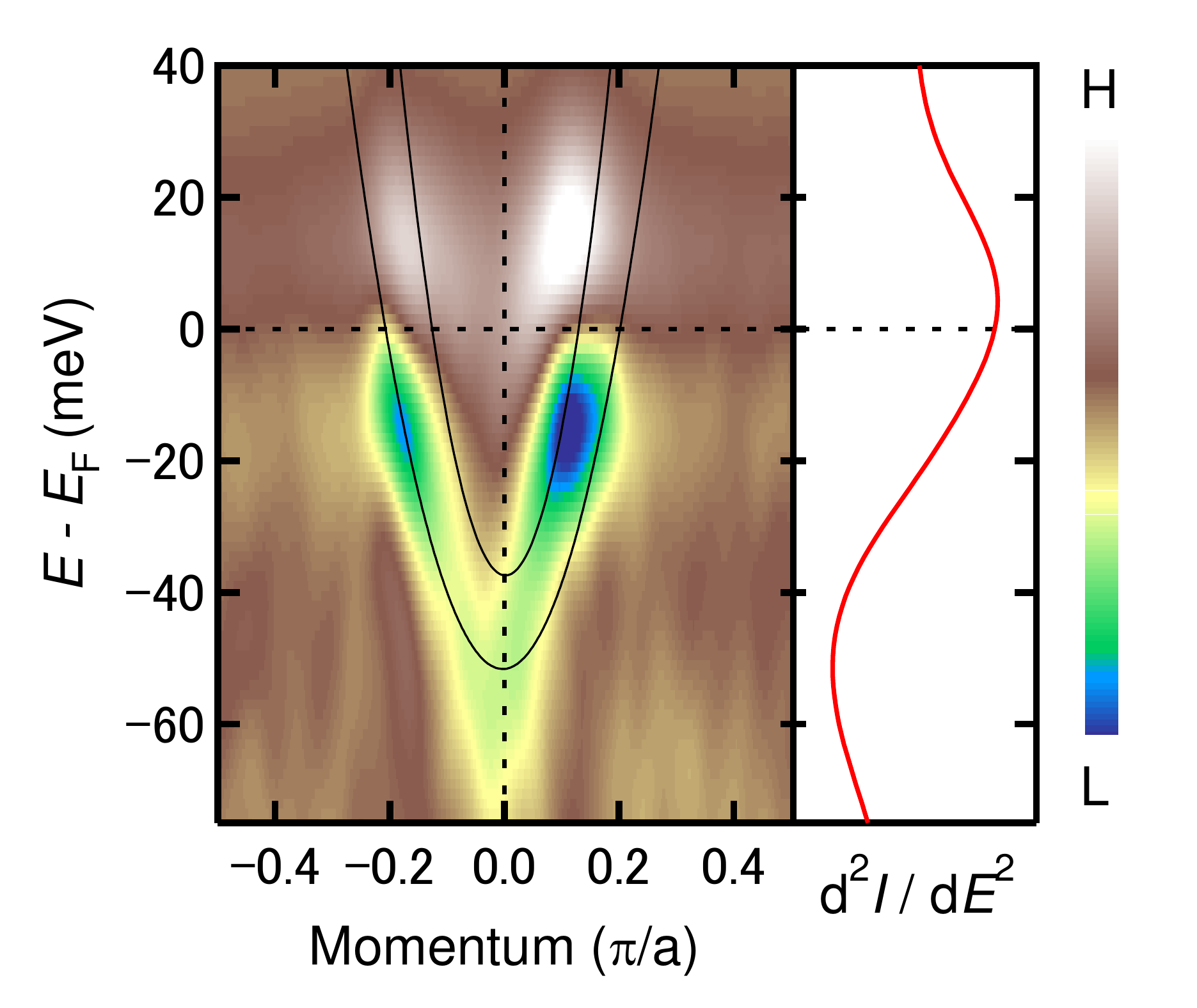}
  \caption{(Color online) Second derivative of the intensity at $X$ point for \Cab. The raw $E-k$ plot is Fig. 1(d) and black curves are same as ones in Fig. 1(d). The red line in the right panel represents $dE^2/d^2I$ at $k = 0$ in the left panel.}
  \label{figS2}
\end{figure}

Figure \ref{figS2} shows the second derivative of the intensity with respect to energy at $X$ point. Although it is hard to see the bottom of the electron band in Fig. 1(d), one can determine that the bottom of the band is located around $E-E_\mathrm{F} \sim 50$ meV from the red line in Fig. \ref{figS2}.

\newpage
\section*{EDCs above and below \Tc\ at $X'$ point}
\begin{figure}[htpb]
  \centering
  \includegraphics[width=120mm,origin=c,keepaspectratio,clip]{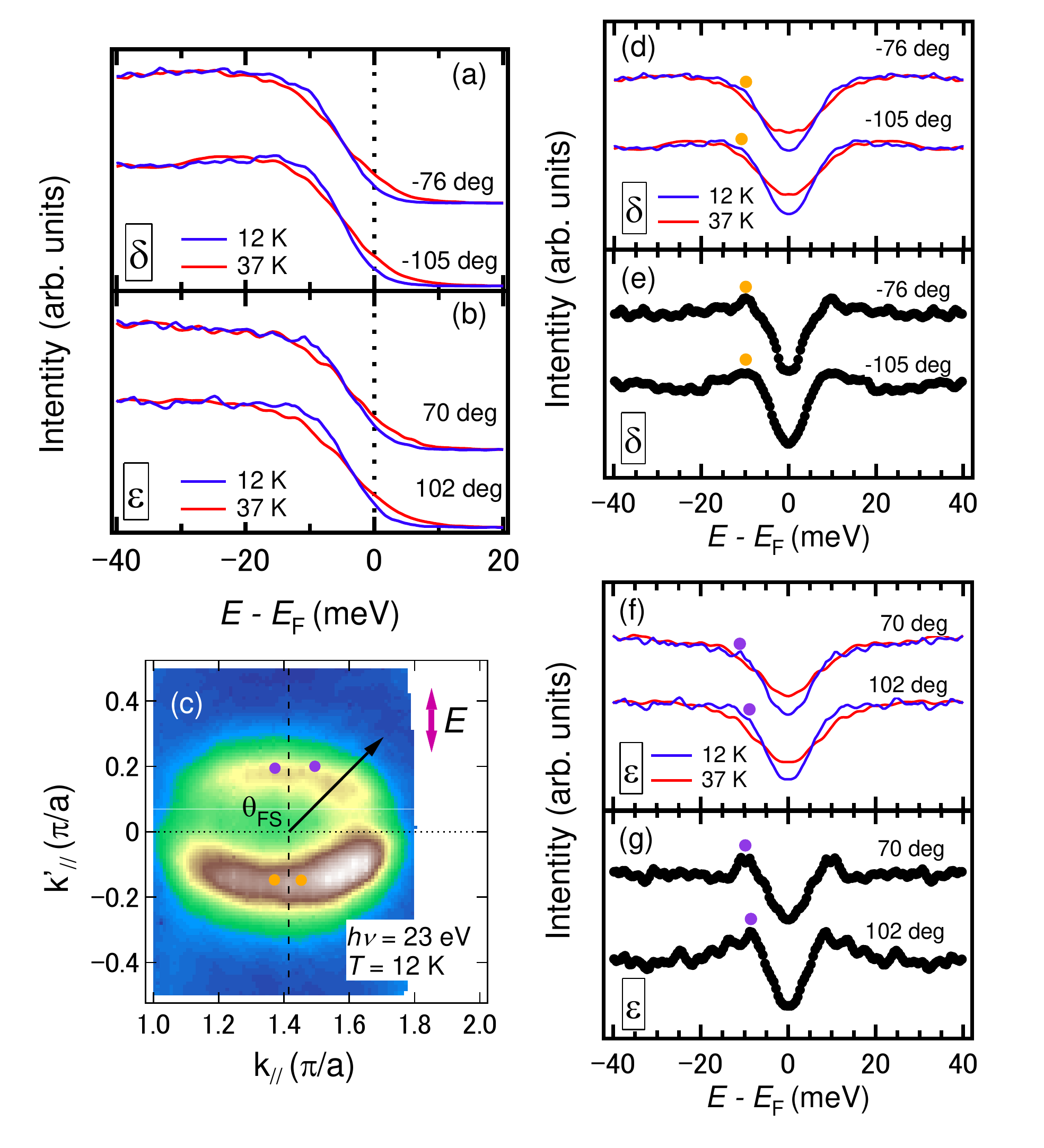}
  \caption{(Color online) (a) and (b) EDCs for \Cab\ around the $X'$ point above and below \Tc. (c) In-plane FS mapping around the $X'$ point. (d) and (f) EDCs symmetrized at \Ef\ for $\delta$ and $\epsilon$ FSs. (e) and (g) Divided spectra I(37K)/I(12K) for $\delta$ and $\epsilon$ FSs. $k_{//}$ and $k'_{//}$ are along $Z$-$X'$ lines.}
  \label{figS3}
\end{figure}

The energy distribution curves (EDCs) for $\delta$ and $\epsilon$ FSs around the $X'$ point above and below \Tc\ are presented in Figs. \ref{figS3}(a) and (b), respectively. In order to see the gap features more clearly, the symmetrized EDCs are shown in Figs. \ref{figS3}(d) and (f). The gaps observed in the EDCs at 37 K are considered to be the pseudogap as mentioned in other studies \cite{Shimojima_P, gap_Y}. Figures \ref{figS3}(e) and (g) show the divided spectra I(37K)/I(12K) where I(37K) is the normal state spectrum and I(12K) is the SC state one. One can see the peak structures that are attributed to the SC gaps. These peak energies are almost the same as the shoulder energies in EDC.

\newpage
\section*{Fitting process of SC gaps}
\begin{figure}[htpb]
  \centering
  \includegraphics[width=80mm,origin=c,keepaspectratio,clip]{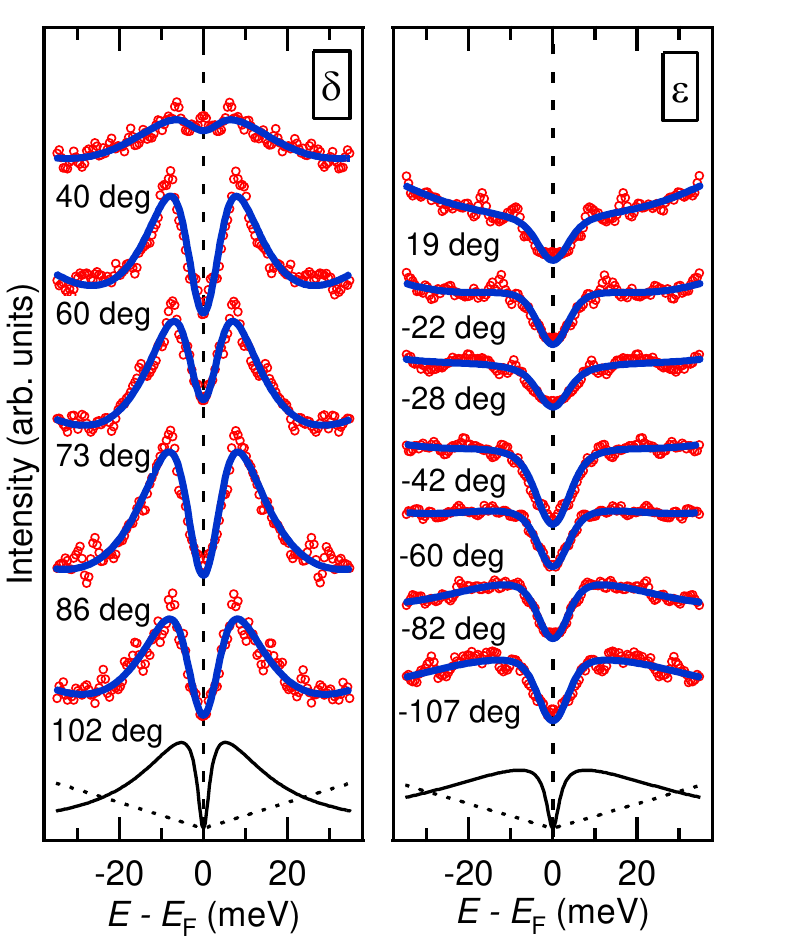}
  \caption{(Color online) Fitting results for \Cab\ around the $X$ point taken at $h\nu = 34$ eV on the $\delta$ and $\epsilon$ FSs. The EDCs for $\delta$ FS are fitted with the phenomenological SC spectral function \cite{Norman2} and a linear background. }
  \label{figS4}
\end{figure}

The all SC gap sizes in the present results were obtained through the fitting process. The symmetrized EDCs for $\delta$ FS around $X$ point are fitted with the phenomenological SC spectral function \cite{Norman2, Zhang}. This spectral function is convoluted by the Gaussian function with the energy resolution = 7 meV and a background is also used (Fig. \ref{figS4}). The fitting results are shown in Fig. \ref{figS4} as an example. 

\begin{thebibliography}{99}
\bibitem{I4mmm} In the first Brillouin zone (BZ) of the space group $I4/mmm$, the $\Gamma$ and the $Z$ points are at ($0, 0, 0$) and ($0, 0, 2\pi/c$), respectively, while the $Z$ point of the adjacent BZ is at ($2\pi/a, 0, 0$). Then, the $\Gamma$ point of the first BZ and the $Z$ point of the adjacent BZ are on the same $k_z = 0$ plane and they share the same electron FSs. Therefore, the shape of the electron FS at ($\pi/a, \pi/a, 0$) is the same as that of the FS at ($\pi/a, \pi/a, 2\pi/c$), if it is rotated by 90 degree.
\end{thebibliography}
\end{document}